# INITIAL BEAM-PROFILING TESTS WITH THE NML PROTOTYPE STATION AT THE FERMILAB A0 PHOTOINJECTOR*

A.H. Lumpkin, R. Flora, A.S. Johnson, J. Ruan, J. Santucci, V. Scarpine, Y.-E. Sun, R. Thurman-Keup, M. Church, and M. Wendt    Fermilab, Batavia, IL U.S.A. 60510

*Abstract*

The beam-profile diagnostics station prototype for the superconducting rf electron linac being constructed at Fermilab at the New Muon Lab has been tested. The station uses intercepting radiation converter screens for the low-power beam mode: either a 100-μm thick YAG:Ce single crystal scintillator or a 1-μm thin Al optical transition radiation (OTR) foil. The screens are oriented with the surface perpendicular to the beam direction. A downstream mirror with its surface at 45 degrees to the beam direction is used to direct the radiation into the optical transport. The optical system has better than 20 (10) μm rms spatial resolution when covering a vertical field of view of 18 (5) mm. The initial tests were performed at the A0 Photoinjector at a beam energy of ~15 MeV and with micropulse charges from 25 to 500 pC for beam sizes of 45 to 250 microns. Example results will be presented.

## INTRODUCTION

Beam-profile diagnostics are being developed for a superconducting radiofrequency (SCRF) Test Accelerator that is currently under construction at the New Muon Lab (NML) at Fermilab [1]. The facility's design goals include the replication of the pulse train proscribed for the International Linear Collider (ILC). An rf photoelectron gun based on the PITZ design will generate the beam. In test-beam mode a low-power beam will be characterized with intercepting radiation converter screens: either a 100-μm thick YAG:Ce single crystal scintillator or a 1-μm thin Al optical transition radiation (OTR) foil. In both cases the screen surface was normal to the beam direction followed by a downstream 45-degree mirror that directed the radiation into the optical system. This configuration does lead to a complication for the OTR imaging since the two sources (the forward OTR from the first foil and the backward OTR from the downstream mirror) are basically equal in strength, but with the second image out of focus. We investigated this latter aspect as part of our studies as well as confirming that the YAG:Ce screen provided beam imaging at the 15-pC charge level. The selected scintillator option provided the screen resolution needed, and the optics also gave us the better than 20-μm rms resolution specification for a field of view of 18-mm diameter.

---
*Work supported by U.S. Department of Energy, Office of Science, Office of High Energy Physics, under Contract No. DE-AC02-06CH11357.

## EXPERIMENTAL BACKGROUND

The tests were performed at the Fermilab A0 photoinjector facility which includes an L-band photocathode (PC) rf gun and a 9-cell SCRF accelerating structure which combine to generate up to 16-MeV electron beams. The drive laser operates at 81.25 MHz although the micropulse structure is counted down to 1 MHz. Due to the low electron-beam energies and OTR signals, we typically summed over micropulses depending on the charge per micopulse. Micropulse charges from 25 to 500 pC were used for beam sigma sizes of 45 to 250 microns. The prototype station was installed in the user beam line section beyond the horizontal spectrometer in the straight ahead line as indicated in Fig. 1.

Based on previous tests in the A0PI beamlines we had identified a configuration to dramatically reduce the magnitude of needed corrections to the simple image sizes that relate to beam size, divergence, or energy spread depending on the diagnostics configuration [2]. The former YAG:Ce 50-μm thick powder screens oriented at 45 degrees to the beam direction (with a spatial resolution term of 60±20μm) were replaced by 100-μm thick YAG:Ce single crystals oriented normal to the beam followed by a 45 degree mirror to direct the radiation to the optical system. This configuration reduces the screen resolution term to less than 10 μm rms based on previous reports [3,4] and also basically eliminates the depth-of-focus issue of the 45 degree scintillator for multiple slit images spread across several mm of the field of view in our emittance-measuring procedures [5-7]. These concepts were applied to the NML prototype station as well.

The prototype station (see Fig. 2) consists of the vacuum cross with a three-position pneumatic actuator allowing selection of a beam impedance matching screen, a 100-μm thick YAG:Ce single crystal with its surface normal to the beam direction followed by a 45 degree turning mirror, or a 1-μm thick Al foil for OTR followed by a 45 degree turning mirror. Initially, both turning mirrors were an aluminized Si substrate (200 μm thick). This is schematically shown in the center of Fig. 3. The basic option is at the left and other possible configurations are shown at the right for other OTR converters or thin Al mirrors. As part of the optics design, a back-illuminated virtual target option with matched field lens could be selected by inserting a beam splitter into the relay optics path. This scene was then relayed to the final Computar zoom lens mounted on the 1.3 Megapixel Prosilica CCD camera and used for resolution and optics calibration aspects. A filter wheel was used to select neutral density filters or one of the two linear polarizers. This prototype

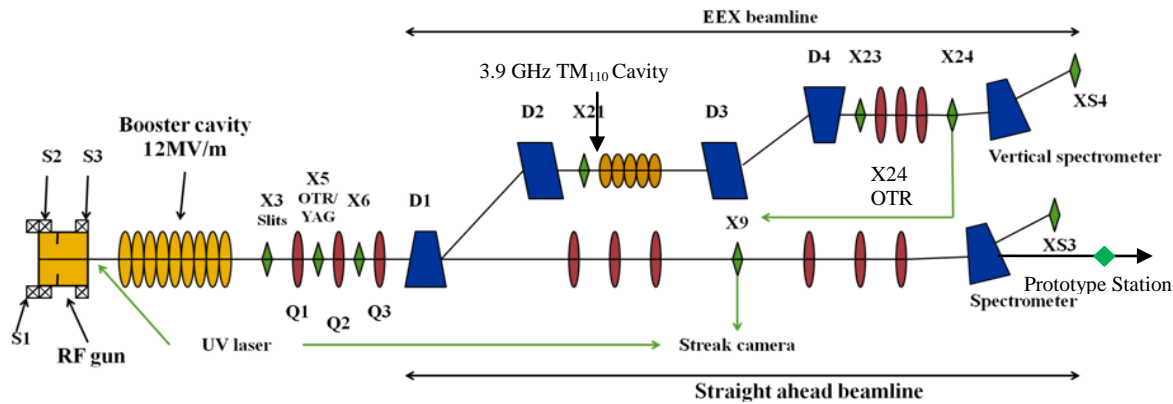

Figure 1: A schematic of the A0 photoinjector test area showing the PC rf gun, 9-cell booster cavity, transverse emittance stations, the OTR stations, the streak camera, and the location of the prototype station.

station was constructed by RadiaBeam Technologies under a contract with Fermilab.

The optical resolution tests were done by using the resolution chart embedded in the virtual target. The modulation transfer function (MTF) and edge resolution function (ERF) were measured for several of the line pair pattern options with the system. Using a Gaussian function convolved with an ideal square line profile, a sigma resolution was determined. The results are summarized in Table I. The optical system has 14 (7) μm rms spatial resolution when covering a vertical field of view (FOV) of 18 (5) mm. The calibration factors were 18.2 μm per pixel and 5.4 μm per pixel, respectively.

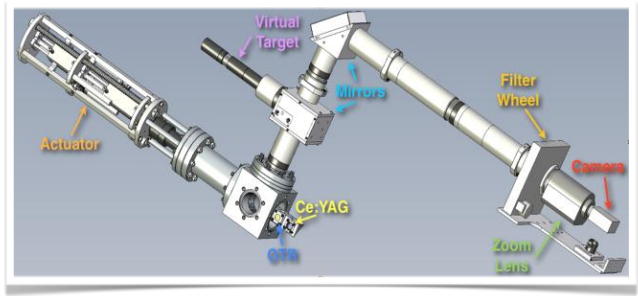

Figure 2: The prototype station showing the double actuator, vacuum cross, OTR and YAG:Ce screens, virtual target, transport optics, filter wheel, zoom lens, and camera. (Drawing courtesy of RadiaBeam Techn.)

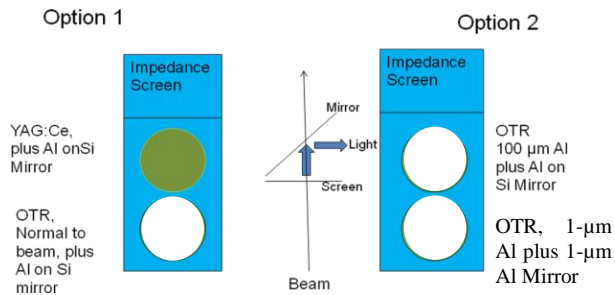

Figure 3: Schematic of the converter screen options considered for the prototype tests. Option 1 was the initial case tested and reported here.

Table I. Summary of the optical resolution test results using the USAF (1951) line patterns on the virtual target.

| Zoom [FOV mm] | Line density [lines/mm] | Modulation [%] | Resolution [μm] |
|---|---|---|---|
| 18 | 16 | 48 | 14 |
| 5 | 29 | 50 | 7 |

## EXPERIMENTAL RESULTS

Typical full image sizes were about 1 mm, but by focusing in one plane with the upstream quadrupoles images can be smaller than 100 μm in one dimension. These beam spots were detected by both the YAG:Ce screen and the OTR screens as summarized below.

### YAG:Ce Scintillator Tests

The first YAG:Ce example in Fig. 4a shows the image for 5 bunches at 500 pC per bunch with an ND1 filter to attenuate the light. The measured projected profile sigma width is 160 μm in the region of interest (ROI). The profile was fit well with a single-Gaussian peak. By using a flat beam transform and irising the drive laser we were able to make a 46- by 295-μm beam image involving only 15 pC of charge in one bunch as shown at the right in Fig. 4b. This demonstrates our low-charge imaging capability.

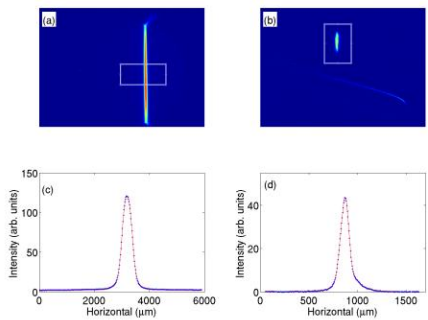

Figure 4: Composite display of a) the vertical band image from the YAG:Ce crystal and c) projected x profile (blue) and its Gaussian profile fit (red curve) and b) the image of the 46 μm x 295 μm beam spot with only 15 pC and d) the corresponding projected profile and Gaussian fit.

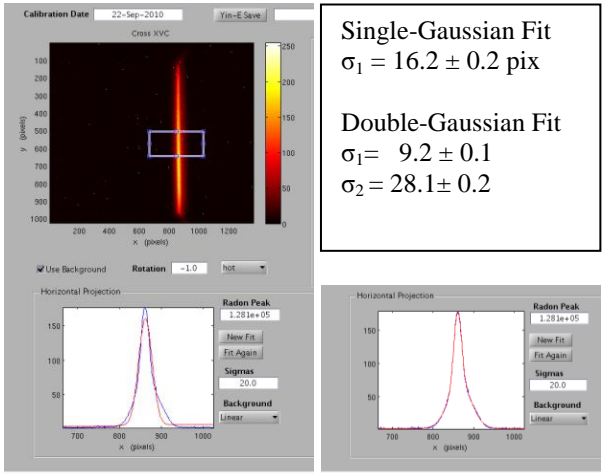 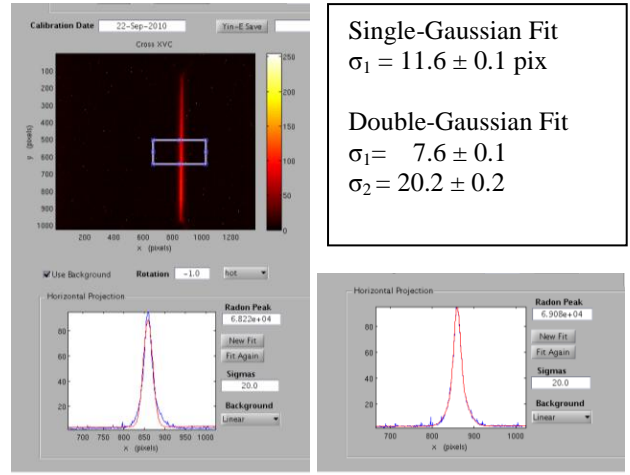

Figure 5: The total radially polarized OTR image (TOP) for a vertical band focus with the single-Gaussian (L) and double-Gaussian fits (R) to the projected profiles.

Figure 6: The vertically polarized OTR image (TOP) for a vertical band focus with the single-Gausssian (L) and two- Gaussian fits (R) to the projected profiles.

### OTR Tests

The OTR configuration was done as a trial for the depth-of-focus issues to see if the second defocused image from the 45-degree mirror could be easily subtracted from the in-focus, first-foil image. However, at these lower gammas we found the second image was about three times larger in size than the focused one. The fitting program initially just did best fits to a single Gaussian or a double Gaussian as seen in Fig. 5 for the total OTR image. The single-Gaussian profile undershoots the data profile shoulders and the peak value, while the double-Gaussian fits the whole profile distribution well as expected since we have two sources at different focuses.

Somewhat unanticipated were the strong effects in the linearly polarized images. We clearly see that the perpendicular polarization component gives smaller sigma fits in Fig. 6 than the total OTR images in Fig. 5. The results are summarized in Table II. The in-focus (narrow) OTR fit of 9.2 pixels compares to the 9.6-pixel result for YAG. In separate tests we have removed the YAG:Ce crystal so that we could directly look at the out-of-focus OTR image from the mirror and compared it to these results. The defocused image of a vertical band's narrow dimension projected profile was generally a Gaussian shape that had a three times wider sigma than the focused one as already indicated in the initial fits of Table II.

Table II. Summary of the Gaussian fit results for the YAG and OTR image profiles including a vertically polarized example.

| Pol. | Ampl. | Sig-x YAG | Sig-x OTR | Ampl. | 2-Gauss. Fit OTR |
|------|-------|-----------|-----------|-------|------------------|
| No   | 121   | 9.6       |           |       |                  |
| No   | 155   |           | 16.2      | 109   | 9.2              |
|      |       |           |           | 69    | 28.1             |
| Vert | 85    |           | 11.6      | 60    | 7.6              |
|      |       |           |           | 32    | 20.2             |

### SUMMARY

In summary, we have performed initial tests on the prototype station for the NML application by using the A0PI beams at 15 MeV. We met the base specifications for imaging the anticipated beam sizes of 300 µm at 40 MeV and 100 µm at 500 MeV at 3.2 nC. In addition we have established applicability with the scintillators at the 15-pC regime for special studies. We also have options to mitigate the microbunching instability in the compressed bright beams using LYSO:Ce crystals as a replacement of the YAG:Ce and a bandpass filter. Further investigations are planned with a fixed focal length lens, a 5 Megapixel camera, and a LYSO:Ce scintillator screen.

### ACKNOWLEDGEMENTS

The authors acknowledge the beamline vacuum work of W. Muranyi and B. Tennis of Fermilab and discussions with M. Ruelas and A. Murokh of RadiaBeam Techn.

### REFERENCES

[1] Sergei Nagaitsev, "ILC-TA at NML", ILC-TA Workshop at Fermilab, November 2006.
[2] A.H. Lumpkin et al., "Upgrades of Beam Diagnostics in Support of Emittance-Exchange Experiments at the Fermilab A0 Photoinjector", Proc. of FEL10, Malmo, Sweden, JACoW (2010).
[3] B.X. Yang (private communication of x-ray results).
[4] M. Maesaka et al., "Beam Diagnostic System of XFEL/Spring8", Proc. of DIPAC09, JACoW (2009).
[5] C.H. Wang et al., "Slits Measurement of Emittance on TTF", International Conference on Accelerator and Large Experimental Physics Control Systems, Trieste, Italy (1999).
[6] R. Thurman-Keup, Emittance code for A0PI, 2009.
[7] L. Lyons, *Statistics for Nuclear and Particle Physicists* (Cambridge University Press, Cambridge 1986).